\documentclass[a4paper,12pt]{extarticle}
\usepackage[top=.9in, bottom=.9in, left=.8in, right=.8in]{geometry}
\usepackage[pagebackref=false,linktoc=page,colorlinks=true,linkcolor=blue,citecolor=blue]{hyperref}

\usepackage[round]{natbib} 

\usepackage{hyperref} \hypersetup{ breaklinks=true, citecolor=blue,
urlcolor=cyan, colorlinks=true, }
\usepackage{scalerel}
\usepackage{algorithm}
\usepackage{algorithmic}
\usepackage{bbm}
\usepackage{geometry}
\usepackage{mathtools}
\usepackage{microtype}
\usepackage{tgtermes}
\usepackage{tgheros}
\usepackage{tgcursor}
\usepackage{multicol}

\usepackage{setspace}
\usepackage{epsfig}
\usepackage{psfrag}
\usepackage[retainorgcmds]{IEEEtrantools}
\usepackage{color}
\usepackage{authblk}
\usepackage{latexsym}
\usepackage{amsmath,amssymb,amsfonts,amsthm}
\usepackage{mathrsfs}
\usepackage{bm}
\usepackage[T1]{fontenc}
\DeclareMathAlphabet\mathbfcal{OMS}{mdugm}{b}{n}
\DeclareMathAlphabet\EuScript{U}{eus}{m}{n}
\SetMathAlphabet\EuScript{bold}{U}{eus}{b}{n}
\usepackage{mathrsfs}
\usepackage{graphicx}
\usepackage{natbib}
\usepackage{pstool,psfrag}
\usepackage{enumitem}
\usepackage[all]{xy}
\usepackage[section]{placeins}

\makeatletter 
\newcommand{\RR}{\ensuremath{\mathbb{R}}}

\newcommand{\NN}{\ensuremath{\mathbb{N}}}

\title{\Large \textbf{A spatial Poisson hurdle model with application to wildfires}}
\author[]{\textbf{Justin A. Kasin}}
\author[]{\textbf{Ioannis Papastathopoulos\footnote{Corresponding author}}}
\affil[]{\small School of Mathematics and Maxwell Institute, University of Edinburgh, Edinburgh, EH9 3FD}
\affil[]{\small  j.kasin@ed.ac.uk \qquad i.papastathopoulos@ed.ac.uk}
 \date{}
\begin{document}

\maketitle
\vspace{-40pt}
\begin{abstract}
  Modelling wildfire occurrences is important for disaster management
  including prevention, detection and suppression of large
  catastrophic events. We present a spatial Poisson hurdle model for
  exploring geographical variation of monthly counts of wildfire
  occurrences and apply it to Indonesia and Australia.\ The model
  includes two a priori independent spatially structured latent
  effects that account for residual spatial variation in the
  probability of wildfire occurrence, and the positive count rate
  given an occurrence. Inference is provided by empirical Bayes using
  the Laplace approximation to the marginal posterior which provides
  fast inference for latent Gaussian models with sparse structures.\
  In both cases, our model matched several empirically known facts
  about wildfires.\ We conclude that elevation, percentage tree cover,
  relative humidity, surface temperature, and the interaction between
  humidity and temperature to be important predictors of monthly
  counts of wildfire occurrences. \color{black} Further, our findings
  show opposing effects for surface temperature and its interaction
  with relative humidity.\color{black}
\end{abstract}
\section{Introduction}
Early wildfire detection is an important problem of considerable
public interest on at least two different fronts.\ From a public
health perspective, wildfires emit toxic fumes and particulates such
as ozone ($\text{O}_3$) and fine particulate matter
($\text{PM}_{2.5}$), which have been estimated to contribute towards a
10,800 annual increase in adult cardiovascular mortality between 1997
and 2006 in Southeast Asia \citep{marlier2013nino}.\ From a
conservationist perspective, one only needs to look at the recently
concluded 2019-20 Australian bushfire season, which has killed more
than one billion animals across 10 million hectares of land
\citep{dickman_2020, lewis2020deathly}, to understand the negative
impact (unmitigated) wildfire occurrences can have on the
environment.\ Currently used detection methods include the use of
strategically placed physical flame and smoke sensors, which are
inexpensive but require a high degree of manual intervention and are
not easily scalable \citep{byousooy12}.\ More advanced sensors
incorporate the use of computer vision algorithms to eliminate the
need for human intervention \citep{mahmoud2018forest,
  bu2019intelligent}, but again require sensors to be deployed in
areas of interest, and is therefore not easily scalable.\ On the other
hand, the availability of free satellite data in recent years has made
remote sensing technology to be increasingly seen a viable alternative
for a wide range of disaster management, ranging from cyclones
\citep{hoque2017tropical} to oil spills \citep{jha2008advances}.

\begin{figure}[htbp!] \centering
  \includegraphics[scale=0.6, trim=10 10 10 10]{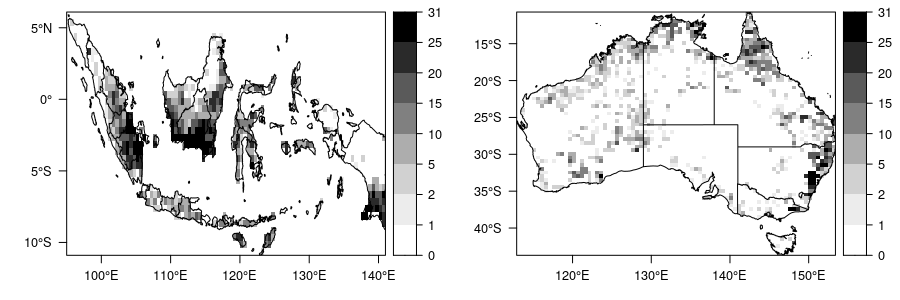}
  \caption{Counts of Indonesian wildfire occurrences in September 2015
    (left) and Australian wildfire occurrences in December 2019
    (right).}
  \label{fig:frpfire_aus_dec_19}
\end{figure}

Typically, wildfire count data come in excess of zeroes.\ This
sparsity is best illustrated by Figure \ref{fig:frpfire_aus_dec_19},
which shows the number of wildfire occurrences in Australia over
December 2019.\ Even as Australia was grappling with one of its worst
bushfire seasons, the map is largely white, indicating that there were
no fires occurring in those areas at all throughout this month.\ On
the flip side, we can clearly see the existence of hot-spots at the
bottom-right of Australia's map in Figure
\ref{fig:frpfire_aus_dec_19}, indicating that when fires do occur,
they do occur quite frequently.

Such characteristics manifest themselves in surprisingly a diverse
range of real world datasets, from insurance claims
\citep{doi:10.1080/10920277.2007.10597487, rohrbeck2018extreme} to
emergency department visits
\citep{doi:10.1111/j.1467-985X.2012.01039.x}.\ How might we begin to
fit a statistical model to such a dataset? Na\"ively applying Poisson
regression is bound to fail because there is an excess of zeros in the
dataset that the model will not be able to account for.\ The nature of
the wildfire data thus motivates the use of a Poisson hurdle model,
which separates the generation mechanism for zero and positive
counts.\ The model comprises two sequential stages; a Bernoulli
distribution first models the binary outcome of whether a wildfire
will occur at a spot.\ If a wildfire does occur, then the hurdle is
crossed, and a separate, truncated-at-zero count data model controls
the distribution of wildfire counts conditioned on a wildfire
occurring at that
spot. 
We provide an application of the Poisson hurdle model with latent
spatial effects to account for residual spatial variation by analyzing
the monthly counts of wildfire occurrences in Indonesia and Australia
in September 2015 and December 2019 respectively.\ The time periods
were selected in a manner that coincided with some of the worst
wildfire season each country has seen; \cite{huijnen2016fire}
estimated fire carbon emissions over Southeast Asia in 2015 to be the
largest since 1997.\ Meanwhile, more than one billion animals have
perished across some 10 million hectares of land during the 2019-20
Australian bushfire season \citep{lewis2020deathly}.\ Overall, our
analysis has shown the effectiveness of using spatial Poisson hurdle
models to tackle count data problems that exhibit (a) excess zeros,
and (b) spatial auto-correlation---the approach is generic enough to be
used in domains outside monthly counts of wildfire occurrences, and
the freely available source code allows interested parties to easily
adapt and tailor our approach to their problem.

The paper is structured as follows.\ Section \ref{sec:poisson_hurdle}
describes the Poisson hurdle generalized linear model with latent
spatial effects that is used throughout the paper.\ Section
\ref{sec:inference} describes inference via marginal posterior
analysis and the Laplace approximation and further, outlines the model
checking tools that are used to validate the spatial Poisson hurdle
model.\ Section \ref{sec:case_study} then puts our work in context and
details how this model can be used to model monthly counts of wildfire
occurrences in Indonesia and Australia.
\section{Statistical modelling}
\label{sec:poisson_hurdle}
\subsection{Covariate modelling with generalized linear model}
We assume a rectangular grid is placed over a broad geographical
region and wildfire counts are made of the individual fire occurrences
in each cell of the grid.\ Let
$\smash{\bm Y = \left(Y_1, \dots, Y_n\right)^{\top} \in \mathbb{N}^d}$
be a random vector where each $Y_i$ represents the total number of of
wildfires in the $i$th cell over a specified period of time.\
Alongside $\bm Y\!$, a matrix of covariates
$\smash{(\bm Z_1,\ldots,\bm Z_n)^\top\in \RR^{n\times k}}$ is also
given.\ We facilitate modelling of covariates via a generalized linear
model (GLM) with Poisson hurdle probability mass function.\ The model
is based on the assumption that when all covariates are held fixed,
the generation mechanism for zero counts is a Bernoulli trial with
covariate-dependent probability of success, whilst the generation
mechanism for positive counts follow a zero-truncated Poisson
distribution with covariate-dependent rate parameter.\ Writing
$\pi(x)=\{1+\exp(-x)\}^{-1}$ and $\lambda(x) = \exp(x)$, $x\in \RR$,
for the canonical link functions of the logistic and log-linear part
of the model respectively, the Poisson hurdle GLM is specified
according to its conditional probability mass function as
\begin{equation}
  \label{eqn:hurdle_covar} \Pr(Y_i = y \mid \bm Z_i = \bm z) =
  \begin{dcases} 1 - \pi(\bm z^\top\bm \beta_0) & y = 0 \\ \pi(\bm
z^\top\bm \beta_0) \frac{\exp \big \{-\lambda(\bm z^\top\bm
\beta_+)\big \} \, {\lambda(\bm z^\top\bm
\beta_+)}^{y}}{y!\,\left[1-\exp\{-\lambda(\bm z^\top\bm
\beta_+)\}\right]} & y \in \NN_{\setminus 0},
\end{dcases}
\end{equation} \sloppy where $\NN_{\setminus 0}=\{1,2,\dots\}$, and
$\smash{\bm \beta_0 = (\beta_{00}, \dots, \beta_{0k})^\top \in
\RR^{k+1}}$ and $\smash{\bm \beta_+ = (\beta_{+0}, \dots,
\beta_{+k})^\top\in \RR^{k+1}}$ are vectors of parameters describing
the effect of covariates for the zero and hurdle part of the model
respectively.\ The maximum likelihood estimator $(\hat{\bm \beta}_0,
\hat{ \bm \beta}_+)$ is obtained by solving a system of $2(k+1)$
equations.\ Since there is no analytic form for the estimator,
numerical methods such as the BFGS iterative method, are typically
used to obtain values for $\hat{\bm \theta}$ \citep{pscl_hurdle}.\ In
a Bayesian setting, inference proceeds by specifying a prior
distribution over $(\bm \beta_0, \bm \beta_+)$.

To analyze monthly counts of wildfire occurrences in Indonesia on
September 2015 and Australia on December 2019, we use the following
covariates
\begin{equation}
  \label{eqn:case_study_covars} \bm z_i^\top\bm \beta_0 =
\beta_{0,\text{intercept}} + \beta_{0,\text{temp}}\,\text{temp}_i +
\beta_{0,\text{RH}}\,\text{RH}_i +
\beta_{0,\text{elev}}\,\text{elev}_i +
\beta_{0,\text{PTC}}\,\text{PTC}_i + \beta_{0,\text{temp}
\,\text{RH}}\text{temp}_i \,\text{RH}_i,
\end{equation} where temp denotes surface temperature, elev denotes
elevation above sea level, and $\text{temp}_i \,\text{RH}_i$ is an
interaction term between surface temperature and relative humidity.\
An interaction term is included to model how the effect of temperature
is moderated by humidity; for example, fires rarely occur in areas
where both temperature and humidity is high
\citep{flannigan1988study}.\ We also use the same set of covariates in
the count part of the model.

This model permits inclusion of covariates in the modelling process in
a simple manner, however, this setup may be viewed as a starting point
than a complete solution to modelling geographical variation in
wildfire counts.\ To account for unobserved factors influencing the
environmental process we incorporate latent spatial effects, that is,
we include spatially dependent random variables (unobserved) in the
model, with the variables being part of the inference.\ This allows us
to account for residual spatial variation both in probability of
occurrence and positive count rate, leading to a model that
accommodates for over-dispersion and spatial clustering in counts.
\subsection{Latent Gaussian processes}
\label{sec:car_model} Let
$\bm U_0 = (U_{01}, \dots, U_{0n})^\top \in \RR^n$ and
$\bm U_+ = (U_{+1}, \dots, U_{+n})^\top\in \RR^n$ be independent
random vectors.\ In what follows, we use a GMRF prior
\citep{10.2307/2984812,rue2005gaussian} which has the form
$\bm U_0 \perp \bm U_+$ with
$\bm U_0\sim \text{MVN}(\bm \mu_0, Q_{0}(\bm \theta_0)^{-1})$ and
$\bm U_+\sim \text{MVN}(\bm \mu_+, Q_{+}(\bm \theta_+)^{-1})$, where
$\text{MVN}(\bm \mu, \bm \Sigma)$ is the multivariate normal
distribution with mean vector $\bm \mu$ and variance-covariance matrix
$\bm \Sigma$.\ The precision matrices are $n\times n$ positive
definite matrices given by
\begin{equation} Q_0(\bm \theta_0) = \tau_0\, (\,\kappa_0^2 \, I_n +
G) \qquad \text{and} \qquad Q_+(\bm \theta_+) =
\,\tau_+\,(\kappa_+^2\,I_n + G),
  \label{eqn:prec_mtrx}
\end{equation} with $\smash{\bm \theta_0 = (\kappa_{0}, \tau_{0}) \in
  \mathbb{R}_+^2}$, $\smash{\bm \theta_+ = (\kappa_{+}, \tau_{+}) \in
  \mathbb{R}_+^2}$, $\bm \mu_0,\bm \mu_+ \in \RR^{n}$, $I_n\in
\RR^{n\times n}$ is the identity matrix and $G\in \RR^{n \times n}$ is
the Laplacian matrix with elements
\begin{equation} G_{ij} = \begin{dcases} d_i & i = j,\\ -1 & j \in
N(i),\\ 0 & \text{otherwise},
  \end{dcases}
  \label{sec:poisson_spatial}
\end{equation}
where $d_i$ denotes the number of
nearest neighbours of cell $i$.\ Equivalently, this model can be
thought of as the latent spatial effect at cell $i$ depending only on
its nearest neighbours.\ Writing $(\bm U_{0})_{\setminus i}$ for $\bm 
U_{0}$ without its $i$th element, we have that the conditional
distribution of $U_{0i}$ given $(\bm U_{0})_{\setminus i}$ satisfies
\begin{equation}
  \label{eqn:cond_dist} U_{0i} \ | \ (\bm U_{0})_{\setminus i} \sim
\mathcal{N}\left( \mu_i + \! \! \! \sum_{j \in N(i)} \! \dfrac{U_{0j}
- \mu_{0j}}{\kappa_0^2 + d_i}, \ \dfrac{1}{(\kappa_0^2 + d_i)\tau_0}
\right),
\end{equation} $\mu_{0j}$ denotes the mean of the latent spatial
effect $U_{0j}$ and $\mathcal{N}(\mu, \sigma^2)$ is the Normal
distribution with mean $\mu \in \RR$ and variance $\sigma^2\in
\RR_+$.\ The conditional distribution of $U_{i+}\mid (\bm
U_+)_{\setminus i}$ is obtained similarly by replacing $0$ to $+$ in
expression \eqref{eqn:cond_dist}.\ The parameters $(\tau_0,\kappa_0)$
and $(\tau_+,\kappa_+)$ are parameters of the precision matrices and
hence describe the covariance of the Gaussian random vectors.\
Informally, they control the scale of the model ($\tau$) and the
inverse strength of correlation ($\kappa$).\ In
what follows, we consider without loss only zero mean Gaussian vectors
so we enforce $\bm \mu_0 = \bm 0$ and $\bm \mu_+ = \bm
0$.\color{black}
\subsection{Spatial Poisson hurdle model} \sloppy The complete
specification of the spatial model is given through the probability
mass functions of the data generating process conditioned on latent
effects and covariates
\begin{equation}
  \label{eqn:hurdle_spatial} \Pr(Y_i = y \mid \bm U_{i}\,;\,\bm
Z_i=\bm z) =
   \begin{dcases} 1 - \pi(\bm z^\top\bm \beta_0 + U_{0i}) & y = 0, \\
\pi(\bm z^\top\bm \beta_0 + U_{0i}) \frac{\exp\{-\lambda(\bm z^\top\bm
\beta_++U_{+i})\}\,{\lambda(\bm z^\top\bm
\beta_++U_{+i})}^{y}}{y!\,\left[1-\exp\{-\lambda(\bm z^\top\bm
\beta_++U_{+i})\}\right]} & y \in \NN_{\setminus 0},
  \end{dcases}
\end{equation} where $\bm U_i=(U_{0i},U_{+i})$.\ This a latent
Gaussian model comprising three sequential layers.\ The first and
outermost layer is the conditionally independent likelihood function
of the data given everything else.\ Given a sample of wildfire counts
$\bm y=(y_1,\ldots, y_n)^\top$ and a matrix of covariates $(\bm
z_{1},\ldots, \bm z_{n})^\top$, the likelihood of $\bm x=(\bm
\beta_0^\top, \bm \beta_+^\top, \bm U_0^\top, \bm
U_+^\top)^\top\in\mathbb{R}^{p}$, $p=2\,(n+k+1),$ is equal to
\begin{equation} L(\bm x\,;\, \bm y, \bm z)= \prod_{i=1}^n \Pr(Y_i=y_i
\mid U_{0i}, U_{+i}\,;\,\bm z).
  \label{eqn:loglik_cov_spatial}
\end{equation} The log-likelihood function is presented in Section
\ref{sec:ll} of the Appendix.\ The second layer comprises the latent
Gaussian field $\bm x$ whose distribution is completely specified the
prior of $(\bm U_0, \bm U_+)$ given in Section~\ref{sec:car_model} and
of $(\bm \beta_0, \bm \beta_+)$.\ For $(\bm \beta_0, \bm \beta_+)$ we
use a $\text{MVN}(\bm 0_{\,2\,(k+1)}, 10^{6}\,I)$ prior, making no
strong assumptions about $\bm \beta_0$ and $\bm \beta_+$.\ Hence the
prior distribution of $\bm x$ is specified as
\begin{equation} \bm x \sim \text{MVN}(\bm 0, Q(\bm \theta)^{-1})\quad
\text{where} \quad Q(\bm \theta) = \begin{bmatrix} 10^{-6}\,I_{k+1} &
0_{k+1, k+1} & 0_{k+1, n} & 0_{k+1, n} \\ 0_{k+1, k+1} & 10^{-6}\,I_{k+1} & 0_{k+1,
n} & 0_{k+1, n} \\ 0_{n, k+1} & 0_{n, k+1} & Q_0(\bm \theta_0) & 0_{n, n} \\
0_{n, k+1} & 0_{n, k+1} & 0_{n, n} & Q_+(\bm \theta_+)
  \end{bmatrix},
  \label{eq:Q}
\end{equation} where $0_{m,n}$ denotes the zero matrix of size
$m\times n$.\ The final layer then consists of a prior distribution
for the hyper-parameter $\bm \theta=(\bm \theta_0, \bm \theta_+)$.\ In
our analysis we used the improper prior $f(\bm \theta)=1$.

Unlike the models specified in Section \ref{sec:poisson_hurdle}, the
introduction of $\bm U_0$ and $\bm U_+$ makes maximum likelihood
unfeasible.\ In what follows, we adopt an empirical Bayes approach
where the parameter vector $\bm \theta = (\bm \theta_0, \bm \theta_+)$
is assumed fixed, and inference then proceeds by maximizing the
marginal posterior.\ The estimation process is described in next
Section.

\section{Inference}
\label{sec:inference}
\subsection{Laplace approximation}
\label{sec:laplace_approx} Let $f(\bm x \mid \bm \theta)$ and $f(\bm
\theta)$ be the prior densities of $\bm x \mid \bm \theta$ and $\bm
\theta$, respectively.\ In this paper we adopt an empirical Bayes
approach where inference proceeds by maximizing an approximation
\citep{tierney1986accurate} to the marginal posterior given by
\begin{equation}
  \label{eqn:marg_post} \widetilde{f}(\bm \theta \mid \bm y) =
\dfrac{f(\bm \theta)\,f(\bm x \mid \bm \theta) \, L(\bm x\,;\, \bm y,
\bm z)}{\widetilde{f}_{G}\,(\bm x \mid \bm \theta, \bm
y)}\,\,\Bigg\vert_{\,{\scaleto{{\bm x = \widehat{\bm x}(\bm
\theta)}\mathstrut}{9pt}}}
\end{equation} where $\widehat{x}(\bm \theta)$ denotes the mode of
$f(\bm x \mid \bm \theta, \bm y)$, $L$ is given in expression
\eqref{eqn:loglik_cov_spatial} and $\widetilde{f}_G(\bm x\mid \bm
\theta, \bm y)$ denotes the probability density function of
$\smash{\text{MVN}(\widehat{\bm x}(\bm \theta), \widehat{\bm
\Sigma}(\bm \theta))}$ where $\smash{\widehat{\bm \Sigma}(\bm
\theta)=\widehat{\bm H}(\bm \theta)^{-1}}$ with
\[ \smash{\widehat{\bm H}(\bm \theta)=-\nabla_{\bm x} \nabla_{\bm
x}^\top \log f(\bm x \mid \bm \theta, \bm y)}\,\vert_{\bm
x=\widehat{\bm x}(\bm \theta)}.
\] and
$\nabla_{\bm x} = (\partial/\partial x_1,\ldots, \partial/\partial
x_p)^\top$.\ The value of $\bm \theta$ that maximizes expression
\eqref{eqn:marg_post} is denoted by $\widehat{\bm \theta}$. The
maximization is done numerically, for example with a gradient-free
method for optimization, see Section \ref{sec:appendix_marg_post} of
the Appendix for more details.\ This entails that for every evaluation
of the approximate marginal posterior \eqref{eqn:marg_post} at a
$\bm \theta$, a search is made in order to find
$\widehat{\bm x}(\bm \theta)$ and $\widehat{ \bm H}(\bm \theta)$.\
This search is typically performed via Newton's method for
optimization \citep{nocedal2006numerical} which starts from an initial
guess $\smash{\bm x^{(0)}\in\RR^p}$ and iterates
\begin{equation} \bm x^{(t+1)} \leftarrow \bm x^{(t)} - \left[
-\nabla_{\bm x}\nabla_{\bm x}^{\top} \log f\{\bm x^{(t)}\mid \bm
\theta, \bm y\} \right]^{-1} \nabla_{\bm x} \log f\{\bm x^{(t)}\mid
\bm \theta, \bm y\}\qquad t=0,1,2,\ldots
  \label{eq:NR}
\end{equation} until convergence, yielding both the mode and the
Hessian at the mode upon termination of the algorithm.\ This method
takes advantage of the sparse structure of precision matrix
\eqref{eq:Q} and is presented in more detail in Section
\ref{sec:Newton} of the Appendix.

Asymptotic $(1-\alpha)$ equal-tail confidence intervals for $x_j$,
$j=1,\ldots, p$ based on asymptotic normality are obtained as
\[\smash{(\hat{x}_{j} + z_{\alpha/2}\,\hat{\sigma}_{x_{j}},
\hat{x}_{j} + z_{1-\alpha/2}\,\hat{\sigma}_{x_{j}})},
\] where $\smash{z_\alpha}$ denotes $\alpha$-th quantile of the
standard normal distribution and $\smash{\hat{\sigma}^2_{x_{j}}}$ is
the diagonal element of \color{black} $\widehat{\bm
\Sigma}(\widehat{\bm \theta})$ \color{black}, corresponding to
$x_{j}$.\ Computing the gradient vector and Hessian matrix of
log-likelihood (\ref{eqn:loglik_cov_spatial}).\ These are derived in
Appendix \ref{sec:gradient_appendix} and Appendix
\ref{sec:hessian_appendix} respectively.
\subsection{Assessing model fit} \color{black} We assess model fit
separately for the logistic and log-linear part of the model using the
receiver operating characteristic (ROC) curve and Pearson residuals
respectively.\ The ROC curve is created by plotting the true positive
rate (TPR) against false positive rate (FPR), where
\begin{equation} \text{TPR}(t) = \dfrac{\sum_{i=1}^{n}
\mathbbm{1}_{\{\Pr(\hat{y}_i > 0) > t, y_i > 0 \}}}{\sum_{i=1}^{n}
\mathbbm{1}_{\{y_i > 0\}}} \text{ and } \text{FPR}(t) =
\dfrac{\sum_{i=1}^{n} \mathbbm{1}_{\{\Pr(\hat{y}_i > 0) > t, y_i = 0
\}}}{\sum_{i=1}^{n} \mathbbm{1}_{\{y_i = 0\}}},
\end{equation} for a range threshold values $t \in [0, 1]$.\ On
average, a model that is not discriminative at all should have its ROC
curve lie perfectly on the dotted diagonal line with an area under
curve (AUC) value of 0.5, while a perfect model would lie perfectly on
the left and top side of the box with an AUC value of 1.

The Pearson residual at cell $i$ is
\begin{equation} r_{i} = \dfrac{y_i - \hat{y}_i}{(\hat{y}_i [ 1 -
\hat{y}_i \exp\{-(\bm z_i^\top\hat{\bm \beta}_+ + \hat{U}_{+,i})
\}])^{1/2}},
\end{equation} with $y_i$ and $\hat{y}_i$ being the actual and
expected monthly count of wildfire occurrences at cell $i$
respectively.  \color{black}

\section{Case study: monthly wildfire counts}
\label{sec:case_study}
\subsection{Monthly wildfires, climate and topographical data} Most
datasets obtained were network common data form (netCDF) files
\citep{ncdf} represented through the \texttt{raster} package
\citep{raster}.\ Due to memory constraints and different standards
used to represent data (e.g.\ different coordinate projections,
different ways of dealing with missing values), occasionally we would
need to preprocess these files using tools such as Climate Data
Operator (CDO) \citep{schulzweida_uwe_2019_3539275}, netCDF operators
(NCO) \citep{nco}, and Geospatial Data Abstraction Library
\citep{gdal}.
\begin{figure}[htbp!] \centering
  \includegraphics[width=\textwidth, trim= 20 20 20 20]{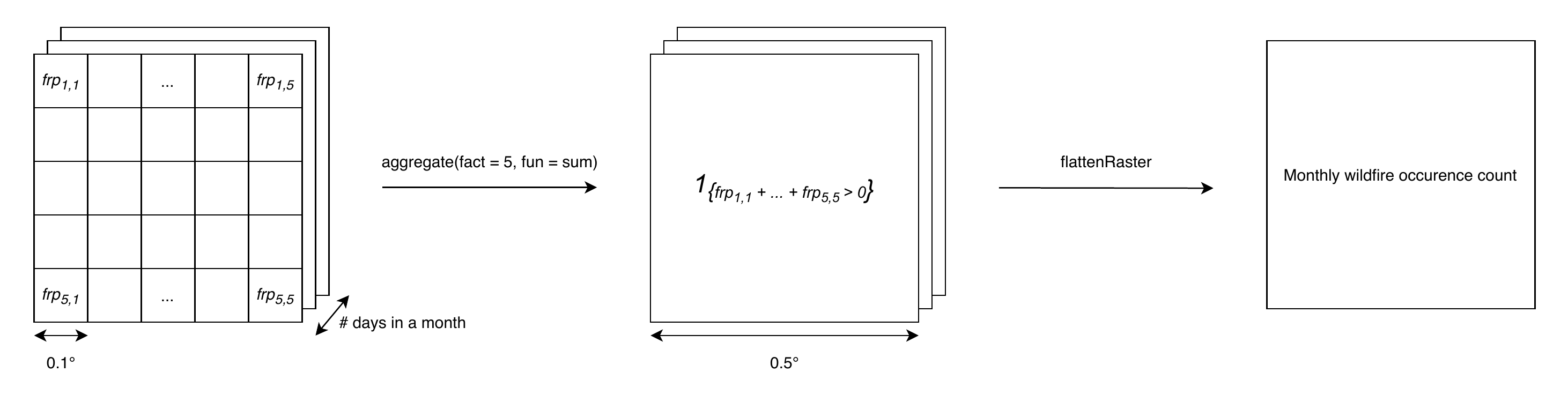}
  \caption{Data preprocessing pipeline for monthly wildfire
    occurrence count data.\ The original daily fire radiative power
    dataset is down-sampled by a factor of 5, binarized, and then
    ``flattened'' across all days in the month.}
  \label{fig:aggregate}
\end{figure} 
Wildfire occurrences data were obtained through the Copernicus
Atmospheric Monitoring Service (CAMS) Global Fire Assimilation System
(GFAS) \citep{gfas}, which provides daily fire radiative power (FRP)
observations from satellite-based sensors, which measures the heat
power emitted by fires \citep{wooster2005retrieval}, with a resolution
of $0.1^{\circ} \times 0.1^{\circ}$ longitude-latitude (lon-lat).\
Near the equator, this corresponds to roughly a grid of size $11
\times 11$ km.
\begin{figure}[htbp!] \centering
  \includegraphics[width=\textwidth]{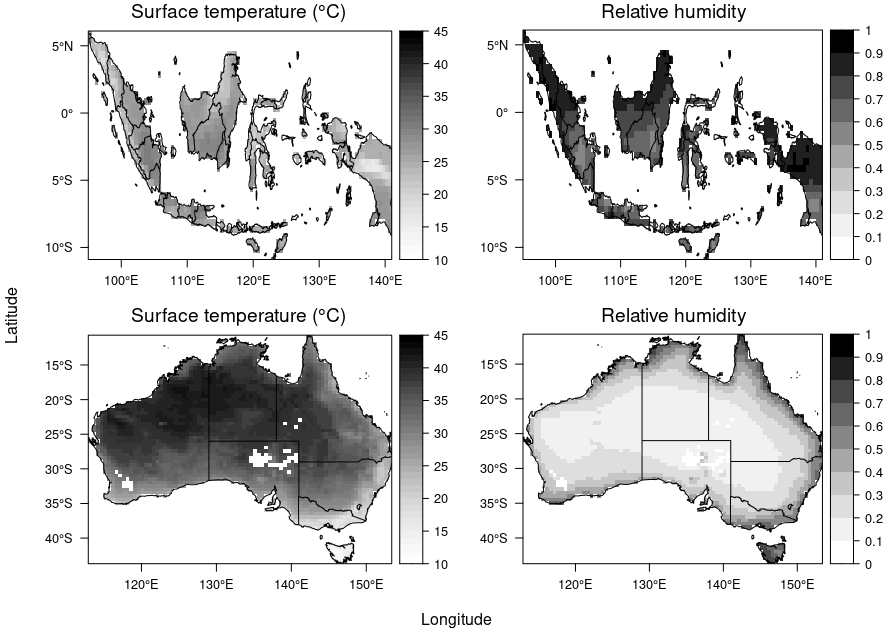}
  \caption{Preprocessed surface temperature (left) and relative
humidity (right) datasets for both Indonesia (top) and Australia
(bottom).}
  \label{fig:climate_vars}
\end{figure}
Due to computational constraints they were down-sampled to a
resolution of $0.5^{\circ} \times 0.5^{\circ}$; the down-sampled grid
is assigned the value of the sum of all the ``component'' grids, as
shown in Figure \ref{fig:aggregate}.\ The data is then converted to be
binary in nature: we assign a cell $i$ in the daily grid with value
$1$ if $frp_s > 0$, and $0$ otherwise.\ As such, we do not
discriminate between fires of differing intensities---they all count
as an occurrence.\ Finally, the binary data is aggregated across all
days in a month, producing a single grid comprising the monthly
wildfire occurrence count data for each cell.

We use surface temperature $(^{\circ}\text{C})$ and relative humidity
as our predictors, both of which have been empirically shown to
influence ignition frequency and therefore the number of wildfire
occurrences \citep{archibald2009limits, jolly2015climate}.\ Surface
temperature dataset was obtained directly through Copernicus Climate
Change Service (C3S) ERA5, an atmospheric reanalyses of the global
climate \citep{era5}.\ The ERA5 reanalyses do not provide relative
humidity out of the box, but can be calculated indirectly using
dew-point temperature $T_{dew}$ and air temperature $T_{air}$ using the
following formula \citep[Equation 7.5]{p16648}:
\begin{equation} \text{RH} =
\dfrac{e_{\text{sat}}(T_{\text{dew}})}{e_{\text{sat}}(T_{\text{air}})},
\qquad\text{with} \qquad e_{\text{sat}}(T) = a_1 \exp{\left\{ a_3
\left( \dfrac{T - T_0}{T - a_4} \right) \right\}},
\end{equation} with $T_0 = 273.16$ K, $a_1 = 611.21$ Pa, \color{black}
$a_3 = 17.502$\color{black}, and $a_4 = 32.19$ K, according to
\cite{buck1981new}.\ The data is then re-sampled to match the
resolution of our wildfire occurrence dataset using bi-linear
interpolation.\ Figure \ref{fig:climate_vars} shows spatial plots of
the resulting datasets.\ Noteworthy here is that there are missing
gaps in the dataset---these are artifacts of our preprocessing
pipeline, any cells with missing value in any of the covariates were
discarded from the analysis (for Indonesia, 3 out of 619 and for
Australia, 49 out of 2788).\
\begin{figure}[htbp!] \centering
\includegraphics[width=\textwidth]{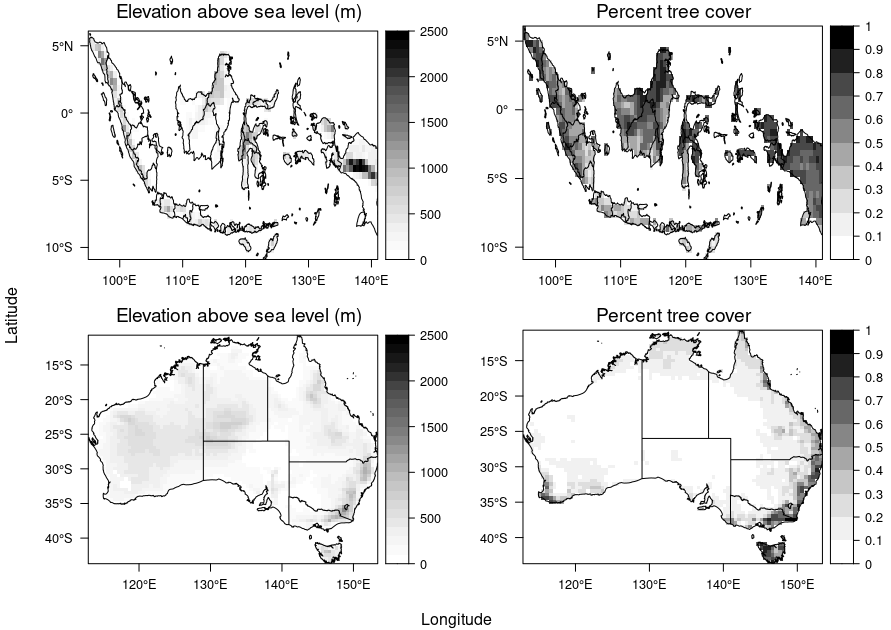}
  \caption{Preprocessed elevation above sea level (left) and percent
tree cover (right) datasets for Indonesia (top) and Australia
(bottom).}
  \label{fig:topographical_vars}
\end{figure}
Topographical variables such as elevation above sea
level (m) have been shown in some cases to be more important than
climate variables \citep{dillon2011both,Serra2014}.\ Moreover,
\cite{archibald2009limits} showed that high percent tree cover
(PTC)---defined as the ratio of area covered by branches and tree
leaves to ground surface when viewed from above---is a good proxy for
high fuel load, thereby allowing fires to propagate more easily than
when PTC is low.\ The datasets were obtained from the Geospatial
Information Authority of Japan \citep{ptc, elevation}, and were
preprocessed in a similar way to climate variables.\ Additionally, we
had to remove values for sea grids using NCO, as the default values
($-9999$ and $254$ for elevation and PTC respectively) caused the
aggregation algorithm to produce wrong values.\ Figure
\ref{fig:topographical_vars} shows spatial plots of the resulting
datasets.

\subsection{Results}
\label{ss:results_ina} Using a tolerance level of
$\epsilon = 1 \times 10^{-7}$, the Nelder--Mead algorithm converged at
$\hat{\bm \theta}_0 = ( 0.390, 0.518 )$ and
$\hat{\bm \theta}_+ = ( 2.565, 0.369 )$ for the dataset of Indonesia.\
As mentioned in Section \ref{sec:car_model}, since $\tau$ and
$\kappa^2$ correspond to the inverse relationship variability and
strength between latent spatial effects respectively, we can interpret
our results as follows. Latent spatial effects for the count process
are more strongly correlated with their neighbours than that for the
hurdle process $(\hat{\kappa}^{-1}_+ > \hat{\kappa}^{-1}_0)$, and the
relationship is tighter $(\hat{\tau}^{-1}_+ > \hat{\tau}^{-1}_0)$.

Table \ref{tab:ina_par} show the value of the estimated coefficients
$\hat{\bm \beta}_0$ and $\hat{\bm \beta}_+$ along with their
approximate $95\%$ confidence intervals based on asymptotic normality.
\begin{table}[ht] \centering
  \begin{tabular}{lcccl} \hline Parameter & $\hat{ \bm \beta}$ & Lower
C.I. & Upper C.I. & Significant? \\ \hline
$\beta_{0,\text{intercept}}$ & 56.385 & 3.638 & 109.131 & \verb+TRUE+
\\ $\beta_{0,\text{temp}}$ & -1.485 & -3.456 & 0.487 & \verb+FALSE+ \\
$\beta_{0,\text{RH}}$ & \color{black}-78.896 & \color{black}-142.910 &
\color{black} -14.882 & \verb+TRUE+ \\ $\beta_{0,\text{elev}}$ &
$2.206 \times 10^{-3}$ & $-4.554 \times 10^{-4}$ & $4.868 \times
10^{-3}$ & \verb+FALSE+ \\ $\beta_{0,\text{PTC}}$ & -2.275 & -5.337 &
0.787 & \verb+FALSE+ \\ $\beta_{0,\text{temp}\cdot\text{RH}}$ & 2.300
& -0.173 & 4.773 & \verb+FALSE+ \\ $\beta_{+,\text{intercept}}$ &
15.432 & 8.763 & 22.100 & \verb+TRUE+ \\ $\beta_{+,\text{temp}}$ &
-0.353 & -0.589 & -0.116 & \verb+TRUE+ \\ $\beta_{+,\text{RH}}$ &
-20.002 & -28.931 & -11.074 & \verb+TRUE+ \\ $\beta_{+,\text{elev}}$ &
$3.468 \times 10^{-4}$ & $-2.371 \times 10^{-4}$ & $9.308 \times
10^{-4}$ & \verb+FALSE+ \\ $\beta_{+,\text{PTC}}$ & -0.490 & -1.023 &
0.043 & \verb+FALSE+ \\ $\beta_{+,\text{temp}\cdot\text{RH}}$ & 0.567
& 0.224 & 0.910 & \verb+TRUE+ \\ \hline
\end{tabular}
\caption{Estimates of $\hat{\bm \beta}_0$ and $\hat{\bm \beta}_+$ when
fitting the model to monthly counts of wildfire occurrences in
Indonesia on September 2015, approximate $95\%$ confidence intervals
based on asymptotic normality and whether they are significantly
different from zero.}
\label{tab:ina_par}
\end{table}

\begin{table}[htbp!] \centering
\begin{tabular}{lcccl} \hline Parameter & $\hat{ \bm \beta}$ & Lower
C.I. & Upper C.I. & Significant? \\ \hline
$\beta_{0,\text{intercept}}$ & -2.715 & -4.428 & -1.002 & \verb+TRUE+
\\ $\beta_{0,\text{temp}}$ & -0.048 & -0.0932 & -0.002 & \verb+TRUE+
\\ $\beta_{0,\text{RH}}$ & 11.214 & -15.126 & -7.302 & \verb+TRUE+ \\
$\beta_{0,\text{elev}}$ & 0.003 & 0.002 & 0.004 & \verb+TRUE+ \\
$\beta_{0,\text{PTC}}$ & 5.279 & 3.789 & 6.770 & \verb+TRUE+ \\
$\beta_{0,\text{temp}\cdot\text{RH}}$ & 0.565 & 0.457 & 0.674 &
\verb+TRUE+ \\ $\beta_{+,\text{intercept}}$ & 1.966 & 0.136 & 3.796 &
\verb+TRUE+ \\ $\beta_{+,\text{temp}}$ & -0.045 & -0.0935 & 0.003 &
\verb+FALSE+ \\ $\beta_{+,\text{RH}}$ & -6.435 & -10.271 & -2.600 &
\verb+TRUE+ \\ $\beta_{+,\text{elev}}$ & $5.679 \times 10^{-4}$ &
$8.380 \times 10^{-5}$ & $1.052 \times 10^{-3}$ & \verb+TRUE+ \\
$\beta_{+,\text{PTC}}$ & 2.254 & 1.433 & 3.075 & \verb+TRUE+ \\
$\beta_{+,\text{temp}\cdot\text{RH}}$ & 0.242 & 0.131 & 0.353 &
\verb+TRUE+ \\ \hline
\end{tabular}
\caption{Estimates of $\hat{\bm \beta}_0$ and $\hat{\bm \beta}_{+}$
when fitting the model to monthly counts of wildfire occurrences in
Australia on December 2019, along with their $95\%$ confidence
intervals and whether they are significantly different from zero.}
\label{tab:aus_par}
\end{table}

From Table \ref{tab:ina_par} we have that all else being fixed, a
$1\%$ increase in relative humidity at a cell is associated with a
$\left( 1 - \exp(- \color{black} 0.789 \color{black}) \right) \times
100\% = 54.57\%$ decrease in the odds of a wildfire occurring at least
once on that cell in September 2015, as we had expected.\ On the other
hand and to our surprise, all other hurdle parameters are not
significant.
\begin{figure}[htbp!] \centering
\includegraphics[width=\textwidth]{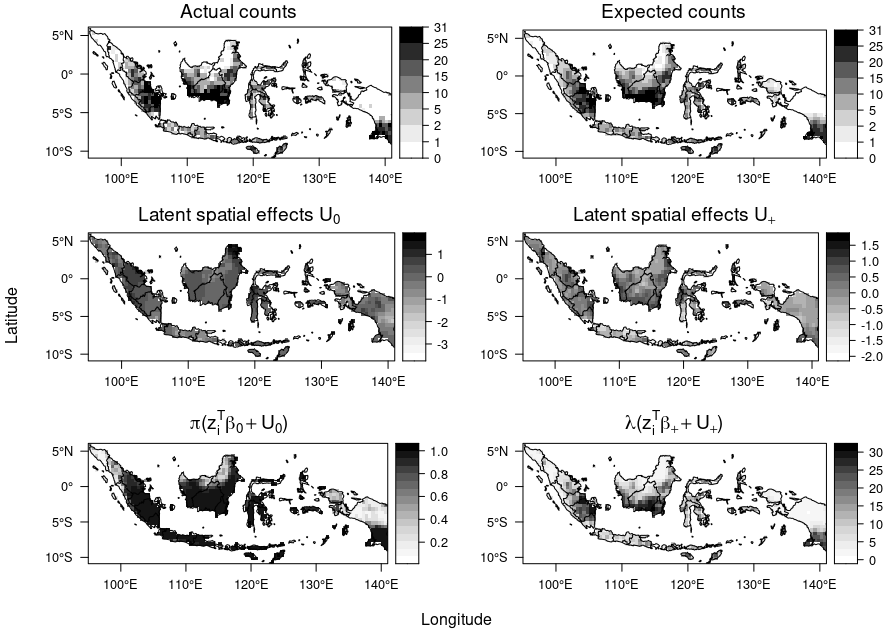}
\caption{Top: Plots of actual (left) and expected (right) counts of
monthly wildfire occurrences.\ Middle: latent spatial effects $\bm
U_0$ (left) and $\bm U_+$ (right).\ Bottom: predicted probabilities of
at least one wildfire occurring $\pi(\bm z_i^\top\bm \beta_0 +
U_{0i})$ (left), and rate parameters $\lambda(\bm z_i^\top\bm \beta_+
+ U_{+i})$ (right) for Indonesia in September 2015.}
    \label{fig:panel_plots_ina}
  \end{figure}
  \begin{figure}[htbp!] \centering
    \includegraphics[width=\textwidth]{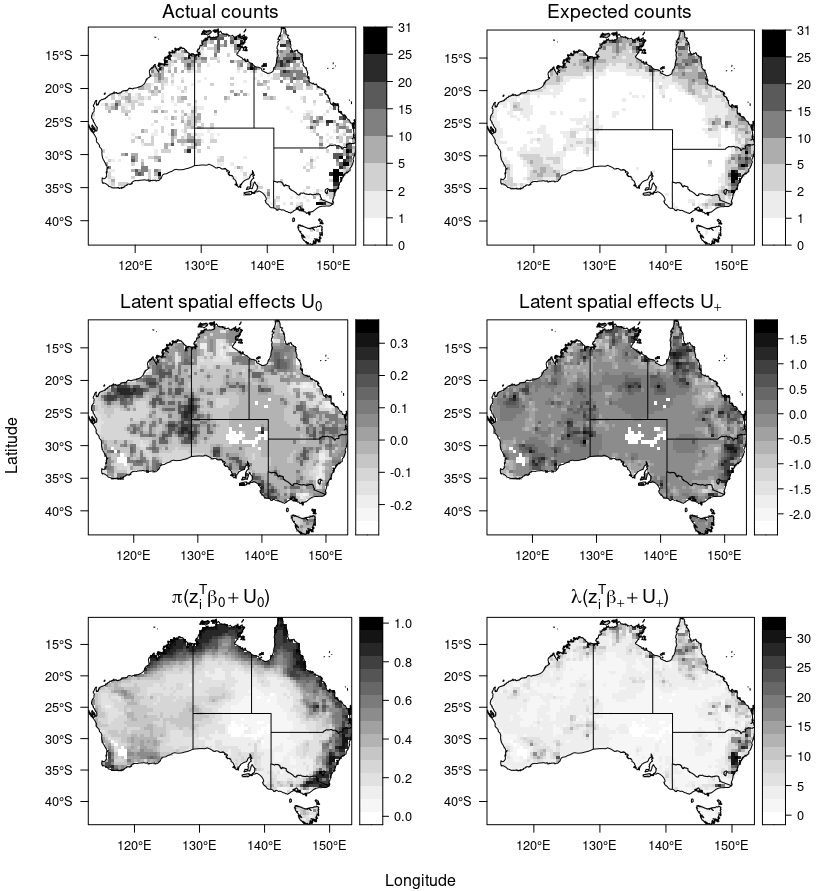}
\caption{Top: plots of actual (left) and expected (right) counts of
monthly wildfire occurrences.\ Middle: latent spatial effects $\bm
U_0$ (left) and $\bm U_+$ (right).\ Bottom: predicted probabilities of
at least one wildfire occurring $\pi(\bm z_i^\top\bm \beta_0 +
U_{0i})$ (left), and rate parameters $\lambda(\bm z_i^\top\bm \beta_+
+ U_{+i})$ (right) for Australia in December 2019.}
    \label{fig:panel_plots_aus}
\end{figure} Regarding the count coefficients, we have that else being
fixed, a $1 \ \text{C}^{\circ}$ increase in surface temperature is
associated with a $29.73\%$ decrease in expected count of wildfire
occurrences on that cell in September 2015, which is surprising as
intuitively we might expect higher surface temperature to be
associated with higher counts of wildfire occurrences.\ The positive
magnitude of the interaction term coefficient cannot account for this
either; we had expected a lower relative humidity to intensify the
effect of temperature on occurrence, but the positive magnitude
implies the contrary.\ This can perhaps be explained by the fact that
we do not actually observe any occurrence where surface temperature is
high and relative humidity is low; from the bottom-right plot Figure
\ref{fig:climate_vars}, we see that the lowest relative humidity
observed in Indonesia is still over $54\%$---any inference involving
values below that is simply extrapolation from our part.\ Figure
\ref{fig:panel_plots_ina} shows that the pattern of the predicted
counts largely matches with those in the actual counts, which is
usually a good sign that the model fit is adequate.\ However, it
appears that our model are predicting small wildfires (light grey)
for cells with no wildfires (dark grey) in the top-left and
top-centre.\ This could be an artifact of our dependency structure
where any cell always depends on its neighbours, forcing the spatial
effects to be smooth all over the region.\ This is also shown in the
middle-left and middle-right plot, which shows that the transition
from regions with low or negative spatial effects to regions with
large spatial effects is gradual.

\begin{figure}[htbp!] \centering
  \includegraphics[scale=0.5]{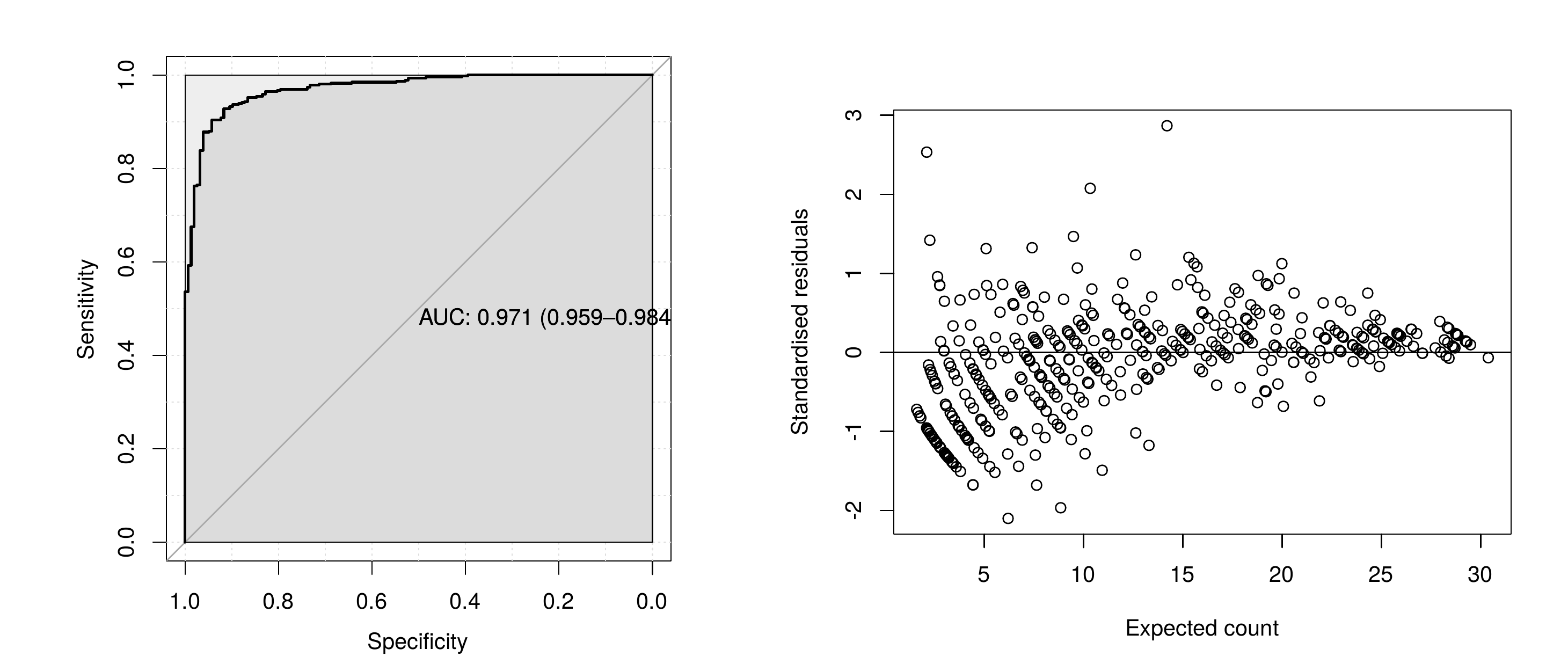}
  \includegraphics[scale=0.5]{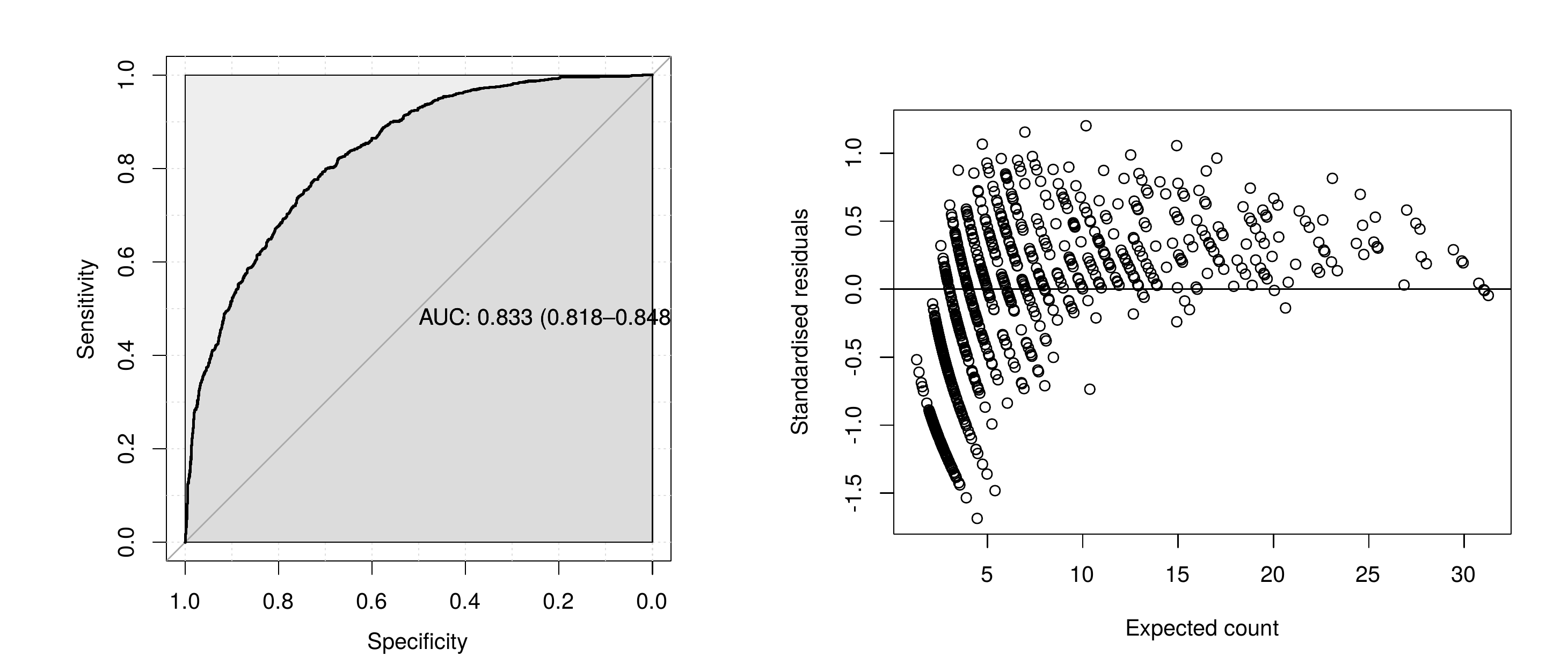}
  \caption{(Left): ROC curve of the hurdle part of the fitted
    Indonesian (top) model and Australian (bottom) model.\ The shape
    of the curve and the high area under curve (AUC) value indicates
    that our model can discriminate between positive and zero cells
    very well.\ (Right): Plot of Pearson residuals against expected
    counts of wildfire occurrences for Indonesian model.\ The
    negative deviations for small expected counts suggest our model
    consistently over-predicts the number of occurrences for
    observations with small actual counts.}
  \label{fig:diagnostics}
\end{figure}

Comparing Figure \ref{fig:climate_vars} with the bottom-left and
bottom-right plots in Figure \ref{fig:panel_plots_ina}, we observe
that the inverse of the pattern in relative humidity are similar to
the pattern in predicted probabilities and rate parameters, which
helps to partially explain the negative sign of the coefficients.\ On
the other hand, it would appear that the relationship between surface
temperature and predicted probabilities and rate parameters are not as
obvious.\ Figure \ref{fig:diagnostics} on the left shows the receiver
operating characteristic (ROC) curve of the fitted model.\ We conclude
that our model can discriminate between cells with at least one
wildfire occurrence and cells with no wildfire occurrence very well,
but this is not surprising at all given that these are in-sample
predictions.\ The high AUC value also confirms our visual inspection
that the red spots in the bottom-left plot of Figure
\ref{fig:panel_plots_ina} match closely with the non-white spots in
the upper-right plot of the same figure.
On the other hand, Figure \ref{fig:diagnostics} on the right shows a
plot of the Pearson residuals against expected counts of wildfire
occurrences.\ The plots show that Pearson residuals for cells with
expected counts of $10$ or less are overwhelmingly negative,
indicating that our model has a tendency to over-predict the number of
wildfire occurrences in cells with few observed counts.\ This also
confirms our observation that the latent spatial effects are smooth
over the entire domain, whereas the actual data comprises more abrupt
boundaries of fire and no-fire zones.\ This is not necessarily a bad
thing, as our model essentially provides a more conservative threshold
for when a zone should be declared as high risk.\ Furthermore, our
main concern are cells with high counts of wildfire occurrences, where
the deviations are not as large.

Due to computational limitations, we used a tolerance level of
$\epsilon = 1 \times 10^{-3}$ in our analysis.\ The Nelder--Mead
converged at $\hat{\bm \theta}_0 = \left( 1.299, 1.326 \right)$ and
$\hat{\bm \theta}_+ = \left( 0.826, 0.481 \right)$ for the dataset of
Australia.\ We can interpret $\hat{\bm \theta}$ in a similar fashion
to what we did in Section \ref{ss:results_ina} to arrive at the same
conclusion, although we must emphasize again that our conclusion is
subject to the reliability of our estimate $\hat{\bm \theta}$.

Table \ref{tab:aus_par} show the value of the estimated coefficients
$\hat{\bm \beta}_0$ and $\hat{\bm \beta}_{+}$ along with their $95\%$
confidence intervals, which can be interpreted in the same manner as
in Section \ref{ss:results_ina}.\ What is noteworthy here is the
differing signs in the coefficients of relative humidity for the
hurdle and count part.\ While higher humidity is associated with lower
counts of wildfire occurrences, as have been empirically shown, it
would appear that higher humidity is associated with a higher
probability of a wildfire occurring at least once.\ This is perhaps
less surprising once we compare the bottom-right plot of Figure
\ref{fig:climate_vars} and the bottom-left plot of Figure
\ref{fig:panel_plots_aus}: in our dataset, wildfire occurrences are
almost guaranteed along the coastal areas, which also happens to be
areas with high relative humidity.\ Without further research one can
only speculate, but a reasonable guess would be that areas with low
relative humidity in Australia tend to be deserts with very few
materials that can act as a fuel for fires to ignite and propagate.\
This is partially supported by the bottom-right plot of Figure
\ref{fig:topographical_vars}, which shows that these areas have low
PTC values.\ It is also interesting to note that surface temperature
and the interaction term show similar effects to the one observed in
Section \ref{ss:results_ina}.\ This perhaps highlight an important
weakness of our climate dataset; using aggregated monthly averages as
our covariates caused us to lose a substantial amount of information
regarding inter-month fluctuations, so our current results could very
well be an exemplification of ecological bias also known as Simpson's
paradox \citep{simpson1951interpretation}, which loosely speaking
highlights the danger of simply relying on aggregate data, which could
mask and obscure actual effects.

Here, the pattern of predicted counts do not match as well as those in
the actual counts.\ \color{black} The smoothness of the latent spatial
effects is even more apparent here\color{black}; in the top-right plot
in Figure \ref{fig:panel_plots_aus} we observe that our model has
essentially linked together three separate clusters of fires in the
north-west, north-east, and south-east, into one smooth curve that
traces the coastline from the top left corner all the way to the
bottom-right corner of Australia. Another interesting observation to
make is that our latent spatial effects $\hat{\bm U}_+$ for the hurdle
part picked up on the wildfire occurrences in the Western Australia,
which our covariates were not able to account for.\ This is shown in
the middle-left plot in Figure \ref{fig:panel_plots_aus}, where there
are black spots in the middle-left area, but not along the coast, since
our covariates can account for those.\ This highlights the advantage
of spatial Poisson hurdle models have over vanilla Poisson hurdle
models described in Section \ref{sec:poisson_hurdle}.

Figure \ref{fig:diagnostics} on the left shows the ROC curve of the
fitted model, and shows that our model is not as discriminative as the
one fitted to the data for Indonesia.\ This largely matches what we
observed in the bottom-left plot of Figure \ref{fig:panel_plots_aus},
and can be attributed to the fact that the covariates were not able to
account for why there are occurrences of wildfires in Central
Australia.\ As a result, our model did not assign a high probability of
at least one observation of wildfire occurrence in these
areas.\ Meanwhile, Figure \ref{fig:diagnostics} on the right show that
as a result of the smoothness of the latent spatial effects, our model
has not only the tendency to over-predict the number of wildfire
occurrences in cells with few observed counts, but also under-predict
the number of wildfire occurrences in cells with large observed
counts.\ This is more worrying, since our model risks providing an
overly optimistic threshold and gloss over what would have actually
been a high risk zone.

Overall, we conclude that globally smooth latent spatial effects may
lack flexibility in modelling geographical variation in monthly counts
of wildfire occurrences.\ As such, potential future work include
exploring methods to constraint the latent spatial effects to not be
globally smooth as for example in \cite{lee2014bayesian} who proposed
a localized conditional auto-regressive prior for modelling residual
spatial auto-correlation, which could serve as a good starting point
for this endeavour.\ Another potential avenue for further research
would be to investigate a latent temporal process for modelling
variation in wildfire occurrences in a single area over time, and
finally combining it with the existing spatial model to create a
spatio-temporal model for modelling counts of wildfire occurrences.

\section*{Acknowledgements}We would like to thank Finn Lindgren for
helpful discussions.
\appendix

\section{Optimization methods}
\subsection{Optimization of approximate marginal posterior}
\label{sec:appendix_marg_post}
To maximize the approximate marginal posterior distribution given by expression
\eqref{eqn:marg_post} we implement the downhill simplex method of
\cite{10.1093/comjnl/7.4.308}.\ In particular, we maximize 
\begin{IEEEeqnarray}{rCl}
  \label{eqn:log_marg_post} \log \widetilde{f}(\bm \theta \mid 
  \bm y) &=& \log f(\bm \theta) + \log
  f(\bm x \mid  \bm \theta) + \log L(\bm x \mid 
  \bm y, \bm z) - \log \widetilde{f}_{G}(\bm x \mid  \bm \theta,
  \bm y).
\end{IEEEeqnarray}
Additionally, the iterative scheme given by \eqref{eq:NR} requires
computation of the gradient and Hessian of the full conditional
density $\log f( \bm x \mid \bm \theta, \bm y )$.\ Here we note that,
by linearity of gradient operator $\nabla_{\bm x}$ and Hessian
operator $\nabla_{\bm x}\nabla_{\bm x}^{\top}$ and the fact that
$f( \bm x \mid \bm \theta, \bm y ) \propto L( \bm x \mid \bm y, \bm z
) \, f( \bm x \ | \ \bm \theta )$, computation of the gradient and
Hessian of the full conditional simply reduces to
\begin{align*} \nabla_{\bm x} \log f( \bm x \mid \bm \theta, \bm y )
  &= \nabla_{\bm x} \log f( \bm y \mid \bm x ) + \nabla_{\bm x}
    \log f( \bm x \mid  \bm \theta ), \text{ and} \\
  \nabla_{\bm x}\nabla_{\bm x}^{\top} \log f( \bm x \mid \bm \theta,
  \bm y ) &= \nabla_{\bm x}\nabla_{\bm x}^{\top} \log f( \bm y \mid
            \bm x ) + \nabla_{\bm x}\nabla_{\bm x}^{\top} \log f( \bm
            x \mid \bm \theta ),
\end{align*} where
$\nabla_{\bm x} \log f( \bm x \mid \bm \theta ) = -Q(\bm \theta)\,(\bm
x - \bm \mu), \quad \nabla_{\bm x}\nabla_{\bm x}^{\top} \log f( \bm x
\mid \bm \theta) = -Q(\bm \theta)$.\ The log-likelihood together with its gradient vector $\nabla_{\bm x} \log L( \bm x \mid \bm y, \bm z )$ and Hessian matrix
$\nabla_{\bm x}\nabla_{\bm x}^{\top} \log L( \bm x \mid \bm y, \bm z )$ are given in Appendix
\ref{sec:ll}, \ref{sec:gradient_appendix} and Appendix \ref{sec:hessian_appendix}
respectively.

\subsection{Log-likelihood}
\label{sec:ll}
The logarithm of the likelihood function $L(\bm x\mid \bm y, \bm z)$
given by expression \eqref{eqn:loglik_cov_spatial} is
\begin{multline}
  \ell(\bm x; \bm y, \bm z) = \sum_{i=1}^{n}\mathbbm{1}_{\{y_i = 0\}}
  \log{\{1 - \pi(\bm z_i^\top \bm \beta_0 + U_{0i})\}} +
  \sum_{i=1}^{n}\mathbbm{1}_{\{y_i > 0\}} \log{\{\pi(\bm z_i^\top \bm
    \beta_0 + U_{0i})\}} \\ + \sum_{i=1}^{n}\mathbbm{1}_{\{y_i > 0\}}
  \,y_i\,(\bm z_i^\top\bm \beta_+ + U_{+i}) -
  \sum_{i=1}^{n}\mathbbm{1}_{\{y_i > 0\}} \lambda(\bm z_i^\top \bm \beta_+ + U_{+i}) \\
  - \sum_{i=1}^{n}\mathbbm{1}_{\{y_i > 0\}} \log{(y_i !)} -
  \sum_{i=1}^{n}\mathbbm{1}_{\{y_i > 0\}} \log{[1-\exp\{-\lambda(\bm
    z_i^\top \bm \beta_+ + U_{+i})\}]}.
  \label{eq:llik_appendix}
\end{multline}
\subsection{Gradient vector of Poisson hurdle log-likelihood}
\label{sec:gradient_appendix}
Routine differentiation of (\ref{eq:llik_appendix}) gives
\begin{align*}
  \dfrac{\partial\ell}{\partial\beta_{0p}} &= \sum_{i=1}^{n} z_{ip}
                                             \left[ \dfrac{\mathbbm{1}_{\{y_i > 0\}}}{1 +
                                             \exp{(\bm z_i^{\top}\bm \beta_0 + U_{0i})}} -
                                             \dfrac{\mathbbm{1}_{\{y_i = 0\}}}{1 + \exp{
                                             \{-(\bm z_i^{\top}\bm \beta_0 + U_{0i})\}}} \right], \\
  \dfrac{\partial\ell}{\partial\beta_{+p}} &= \sum_{i=1}^{n} z_{ip}
                                             \mathbbm{1}_{\{y_i > 0\}} \left[y_i -
                                             \exp(\bm z_i^{\top}\bm \beta_+ + U_{+i}) -
                                             \dfrac{\exp(\bm z_i^{\top}\bm \beta_+ +
                                             U_{+i})}{\exp\{\exp(\bm z_i^{\top}\bm \beta_+ + U_{+i})\} -
                                             1}\right], \\
  \dfrac{\partial\ell}{\partial
  U_{0i}} &= \dfrac{\mathbbm{1}_{\{y_i > 0\}}}{1 +
            \exp{(\bm z_i^{\top}\bm \beta_0 + U_{0i})}} -
            \dfrac{\mathbbm{1}_{\{y_i = 0\}}}{1 + \exp{
            \{-(\bm z_i^{\top}\bm \beta_0 + U_{0i})\}}}, \text{ and} \\
  \dfrac{\partial\ell}{\partial
  U_{+i}} &= \mathbbm{1}_{\{y_i > 0\}} \left\{y_i -
            \exp(\bm z_i^{\top}\bm \beta_+ + U_{+i}) -
            \dfrac{\exp(\bm z_i^{\top}\bm \beta_+ +
            U_{+i})}{\exp\{\exp(\bm z_i^{\top}\bm \beta_+ + U_{+i})\} -
            1}\right\},
\end{align*} where $p = 0, \dots, k$ and $i=1,\ldots,n$.

\subsection{Hessian matrix of Poisson hurdle log-likelihood}
\label{sec:hessian_appendix}
Routine differentiation of (\ref{eq:llik_appendix}) gives
\begin{align*}
  \dfrac{\partial^2 \ell}{\partial\beta_{0q}
  \partial\beta_{0p}} &= - \sum_{i=1}^{n}
                        \dfrac{z_{ip}z_{iq}}{\eta_0^{\prime} + {\eta_0^{\prime}}^{-1} + 2}, \\
  \dfrac{\partial^2 \ell}{\partial\beta_{+q}
  \partial\beta_{+p}} &= \sum_{i=1}^{n} z_{ip}z_{iq}\mathbbm{1}_{\{y_i >
                        0\}}\left[\dfrac{\eta_+^{\prime}\left\{(\eta_+^{\prime} -
                        1)\exp(\eta_+^{\prime}) + 1\right\}}{\left[\exp(\eta_+^{\prime}) -
                        1\right]^2} - \eta_+^{\prime} \right], \\
  \dfrac{\partial^2 \ell}{\partial\beta_{0p}
  \partial U_{0i}} &= -\dfrac{z_{ip}}{\eta_0^{\prime} +
                     {\eta_0^{\prime}}^{-1} + 2}, \\
  \dfrac{\partial^2 \ell}{\partial\beta_{+p}
  \partial U_{+i}} &= z_{ip} \mathbbm{1}_{\{y_i > 0\}}
                     \left[\dfrac{\eta_+^{\prime}\left\{(\eta_+^{\prime} -
                     1)\exp(\eta_+^{\prime}) + 1\right\}}{\left\{\exp(\eta_+^{\prime}) -
                     1\right\}^2} - \eta_+^{\prime} \right], \\
  \dfrac{\partial^2 \ell}{\partial U_{0i}^2}
                      &= -\dfrac{1}{\eta_0^{\prime} + {\eta_0^{\prime}}^{-1} + 2}, \\
  \dfrac{\partial^2 \ell}{\partial U_{+i}^2}
                      &= \mathbbm{1}_{\{y_i > 0\}}
                        \left[\dfrac{\eta_+^{\prime}\left\{(\eta_+^{\prime} -
                        1)\exp(\eta_+^{\prime}) + 1\right\}}{\left\{\exp(\eta_+^{\prime}) -
                        1\right\}^2} - \eta_+^{\prime} \right], \text{ and} \\
  \dfrac{\partial^2 \ell}{\partial\beta_{+q}
  \partial U_{0p}} &= \dfrac{\partial^2 \ell}{\partial\beta_{0q}
                     \partial U_{+p}} = \dfrac{\partial^2 \ell}{\partial\beta_{+q}
                     \partial\beta_{0p}} = \dfrac{\partial^2 \ell}{\partial U_{+j} \partial
                     U_{0i}} = 0,
\end{align*} where $\eta_0^{\prime} =
\exp(\bm z_i^{\top}\bm \beta_0 + U_{0i})$ and
$\eta_+^{\prime} = \exp(\bm z_i^{\top}\bm \beta_+ +
U_{+i})$, $p,q = 0, \dots, k$ and $i=1,\ldots,n$.
\subsection{Newton's method for optimization}
\label{sec:Newton}
\begin{algorithm}
  \begin{algorithmic} \STATE {\bfseries Inputs:} hyper-parameter
$\bm \theta$, response variable $\bm y$, covariates
$\bm z$, tolerance level $\epsilon$; \STATE {\bfseries
Initialize} $\bm x^{(0)} \leftarrow \bm 0$; \STATE
{\bfseries Initialize} step size $\alpha \leftarrow 1$; \FOR{$t = 0,
1, 2, \dots$} \STATE $\bm x^{(t+1)} \leftarrow \bm x^{(t)} -
\alpha  \left[ -\nabla_{\bm x}\nabla_{\bm x}^{\top} \log f\{\bm x^{(0)}\}
\right]^{-1} \nabla_{\bm x} \log f\{\bm x^{(0)}\}$; \WHILE{$\log
f\{\bm x^{(t+1)}\} < \log f\{\bm x^{(t)}\}$} \STATE $\alpha
\leftarrow \alpha/2$; \STATE $\bm x^{(t+1)} \leftarrow
\bm x^{(t)} - \alpha  \left[ -\nabla_{\bm x}\nabla_{\bm x}^{\top} \log
f\{\bm x^{(0)}\} \right]^{-1}  \nabla_{\bm x} \log
f\{\bm x^{(0)}\}$; \IF{$\alpha \leq 2^{-9}$} \STATE {\bfseries
stop} and let $\bm x^{(t+1)} \leftarrow \bm x^{(t)}$; \ENDIF
\ENDWHILE \IF{$\left| \left\Vert \bm x^{(t+1)} \right\Vert -
\left\lVert \bm x^{(t)} \right\lVert \right| < \epsilon$} \STATE
{\bfseries stop} and return $\bm x^{(t+1)}$; \ENDIF \ENDFOR
    \end{algorithmic}
    \caption{Newton's method; for brevity
      $f(\bm x) \equiv f( \bm x \mid \bm \theta, \bm y )$.}
  \label{alg:damped_newton}
\end{algorithm}
{\small
  
\end{document}